\documentclass[global,twocolumn]{svjour}
\usepackage{latexsym,amsmath}
\usepackage{graphics}
\journalname{Applied Physics A}
\begin{document}
\title{Cluster expansion for dimerized spin systems}
\author{H.-J. Mikeska\inst{1}\inst{2} \and  M. M\"uller\inst{1}
}                      % Do not remove
%
%\offprints{M. M\"uller\inst{1}}          % Insert a name of corresponding author
%
\institute{Institut f\"ur Theoretische Physik, Appelstrasse 2, D-30167 Hannover, Germany \and Hahn-Meitner-Institut, Glienicker Strasse 100, D-14109 Berlin, Germany}
\date{Received: date / Revised version: date}
\maketitle
\begin{abstract}
We have studied dimerized spin systems by realizing the cluster
expansion to high order. We have extended our previous dimer
expansion for one-dimen\-sional systems to cover weakly interacting
chains for a quantitative description of three dimensional materials
like (C$_4$H$_{12}$N$_2$)Cu$_2$Cl$_6$ (= PHCC) and KCuCl$_3$. By
comparison with recent inelastic neutron scattering data we are
able to determine the exchange energies between individual
spins.
We have further investigated the incommensurate
region of zigzag chains with isotropic exchange coupling constants
near the disorder-line where the dispersion curve exhibits a minimum
at a finite wave\-vec\-tor. Our approach clearly shows the gradual
transition between the minimum of the dispersion at wave\-vec\-tor $0$ and
wavevector $\pi$ within this region. The extent of the incommensurate
regime is given analytically in an expansion in the coupling
constants.
\end{abstract}
\section{Introduction}
\label{intro}
Spin systems consisting of dimers as basic building blocks describe a
large number of real materials when additional exchange couplings of
smaller magnitude between dimers are considered. For a quantitative
treatment of the elementary excitations of such systems we have
developed the dimer perturbation theory \cite{Bro73,Uhr97} to cover
dimers interacting in two and three spatial dimensions up to high
orders in the interdimer interactions. In this contribution we present
the application of this method to study quantitatively the elementary
excitations in PHCC and KCuCl$_3$ (sect.~\ref{sec:3D}) and to
determine the incommensurate regime of the one-dimensional zig-zag
chain close to the dimer point (sect.~\ref{sec:IC}). The results for
PHCC and KCuCl$_3$ were motivated by and will be compared to spectra
obtained in recent inelastic neutron scattering experiments.

For weakly interacting dimers the cluster expansion is a powerful tool
to calculate power series for properties like groundstate energy
\cite{GSH90} or one magnon excitations \cite{G96}. Due to the large
number of different exchange couplings the obtained power series have
one parameter only which is an overall scaling of the exchange
couplings and finally set to one. Different sets of exchange
parameters require new calculations whereas the dependence in $\vec
q$-space is given completely. The power series which are correct for
the infinite system are build from finite clusters. In our case going
to sixth order the number of topological different colored clusters is
18084 for KCuCl$_3$ and to fourth order 405 for PHCC.

In our notation the basic intradimer interaction (which sets the
energy scale) is denoted by $J$; interdimer interactions are denoted
by $J_{(lmn)}^{(ij)}$ with $(lmn)$ giving the direction between the
dimer centers (in units of the lattice constants) and $(ij)$
specifying the pair of spins which interact with exchange energy
$J_{(lmn)}^{(ij)}$. The $z-$component of the vector from spin $i=1$
to spin $i=2$ on a given dimer is taken to be $> 0$.   

Using this method, it is possible to determine the exchange
interactions between the individual spins forming the dimers whereas
the standard method to analyze systems of interacting dimers is an
RPA-like approach \cite{LSGF84} which allows to determine only the effective
dimer interactions
$J^{\rm eff}_{(lmn)} = (J_{(lmn)}^{(11)} + J_{(lmn)}^{(22)} 
                     - J_{(lmn)}^{(12)} - J_{(lmn)}^{(21)})/2. $ 
This effective approach reproduces the leading terms of the full
series; the full series, however, has additional terms starting in
second order which turn out to be important.

\section{Chains interacting in two and three dimensions: PHCC and KCuCl$_3$}
\label{sec:3D} 
In this section we present our results for the real interacting dimer
compounds (C$_4$H$_{12}$N$_2$)Cu$_2$Cl$_6$ (Piperazinium
Hexachlorodicuprate = PHCC) and KCuCl$_3$. In both these materials
Cu$^{2+}-$ions with $S=\frac{1}{2}$ are responsible for the magnetic 
properties.  

\begin{figure}[ht]
\centering
\resizebox{0.46\textwidth}{0.26\textheight}{\includegraphics{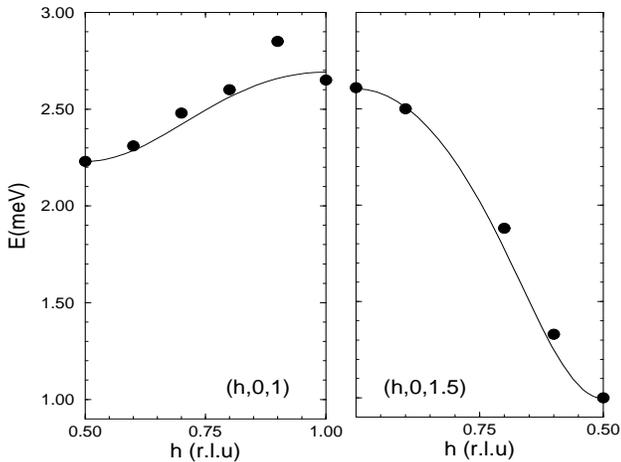}}
%\vspace{3cm} }      % Give the correct figure height in cm
\caption{Dispersion of PHCC for two directions: solid lines denote the result obtained from cluster expansion, circles are data taken from \cite{SZRB01}.}
\label{fig:PHCC}       % Give a unique label
\end{figure}

PHCC was investigated recently by inelastic neutron scattering
\cite{SZRB01} and found to be effectively two-dimensional
(i.e. $J_{(lmn)}^{(ij)}=0$ for $m \ne 0$). The compound can be
considered as built from interacting spin ladders. The inelastic
neutron scattering spectra were analyzed using the effective dimer
model. Satisfactory agreement with the data was obtained when
effective interactions $J_{(lmn)}^{(ij)}$ with $(lmn) = (200)$ and
$(002)$ were introduced as nonzero; however, no obvious exchange paths
exist for these separations. In the following we present an analysis
of the spectra with interactions $J_{(lmn)}^{(ij)}$ between
neighbouring dimers only. The results shown in figure~\ref{fig:PHCC}
are obtained with the set of exchange interactions given in table
\ref{tab:1}. A detailed analysis shows that for PHCC the additional
contributions to the dimer series which are not taken into account by
the effective dimer approach are essential to lower the dispersion
curves. On the other hand we find from a comparison of the complete
and the fourth order result of the effective approach that higher
orders can be safely neglected.

\begin{figure}[ht]
\centering
\resizebox{0.46\textwidth}{0.28\textheight}{\includegraphics{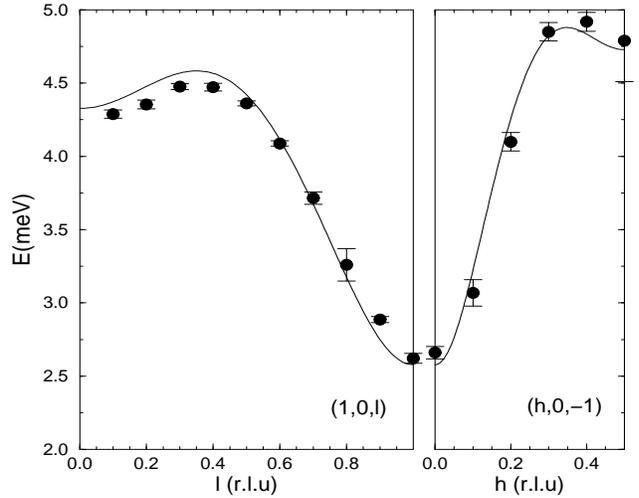}}
%\vspace{3cm} }      % Give the correct figure height in cm
\caption{Dispersion of KCuCl$_3$ for two directions: solid lines denote the result obtained from cluster expansion, circles are data from \cite{CHF99}.}
\label{fig:KCuCl3}       % Give a unique label
\end{figure}

The material KCuCl$_3$ is similar in structure to PHCC. It contains,
however, interdimer interactions also in the remaining (third) spatial
direction. Inelastic neutron scattering experiments on this material
have been published by two groups \cite{KTTS98,CHF99} and have been
interpreted using the effective dimer model. In earlier work
\cite{MM00} we analyzed these data using dimer series expansions to
fourth order. Recently we improved the computer implementation and in
figure~\ref{fig:KCuCl3} we show a spectrum covering all contributions up to
sixth order. This result demonstrates the convergence of our series
expansion for the exchange constants between the individual spins
relevant for KCuCl$_3$ as given in table \ref{tab:1}. It confirms our
previous result that the one-dimensional subunits in this material are
chainlike structures with alternating couplings rather than ladders.
\section{Incommensurate regime in zig-zag chains}
\label{sec:IC} 
\begin{figure}[ht]
\centering
\resizebox{0.46\textwidth}{0.24\textheight}{\includegraphics{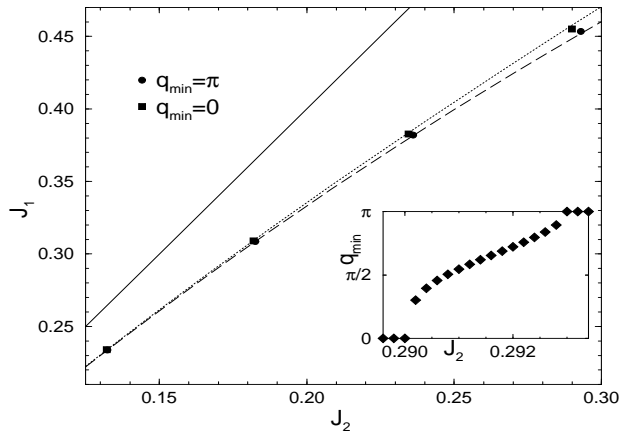}}
	\caption{Incommensurate regime: The solid line denotes the Shastry-Sutherland line, the other lines are eqs.~(\ref{eq:regime}) up to fifth order. The inset shows the wavevector $q_{min}$ for the minimum value of the dispersion on the line $J_1=\frac 1 2 (1.2 - J_2)$. All symbols are obtained from expansions up to twelfth order.}
\label{fig:2}
\end{figure}

In this section we consider a one-dimensional array of dimers forming
an interacting zig-zag chain, $J_{(100)}^{(11)} = J_{(100)}^{(22)} =
J_2, \; J_{(100)}^{(21)} = J_1$.  This system has the interesting
feature of an incommensurate regime where the minimum of the basic
magnon dispersion curve is at finite wavevector $q=q_{min}$, $0 <
q_{min} < \pi$. The incommensurate regime is close to the disorder
line (Shastry-Sutherland line) $J_1 = 2 \; J_2$, where noninteracting
dimers form the exact ground state, but requires slightly larger
values of $J_2$ for given $J_1$. Using the dimer expansion to sixth
order we have determined the limiting lines of the incommensurate
regime as
\begin{gather}
J_{1d}(J_2) \le J_1  \le  J_{1u}(J_2)\nonumber\\
\begin{align}
J_{1d}(J_2) &=2J_2-2J_2^2+2J_2^3-\frac 5 2 J_2^4+\frac{21} 4 J_2^5+\mathcal O(J_2^6)\nonumber\\
J_{1u}(J_2) &=2J_2-2J_2^2+2J_2^3-\frac 1 2 J_2^4+\frac 3 4 J_2^5+\mathcal O(J_2^6)
\label{eq:regime}
\end{align}
\end{gather}
Thus the deviation from the Shastry-Sutherland line is of second order
in the interdimer interaction whereas the width of the incommensurate
regime is of fourth order. The incommensurate regime is shown in
figure~\ref{fig:2} in a comparison of the results from high order expansions
(which agree with the findings from the diagonalization of finite
systems using the Lanczos algorithm) and the approximation of
eqs.~(\ref{eq:regime}). The behaviour of the wave vector $q_{min}$,
corresponding to the minimum excitation frequency is also shown for
one path through the incommensurate regime in twelfth order (thus
improving on the results shown in ref.~\cite{MM00}).

Evidently, the incommensurate regime is tiny and therefore so far
mostly of academic interest; however, if a material becomes available
which allows tuning the exchange parameters by e.g. alloying or
applying external pressure, our results provide a quantitative
guide where to expect this regime of considerable qualitative
interest.

\section{Conclusions}
\label{sec:concl}
Using the dimer expansion after computer implementation we have
investigated the low energy dynamics of the real materials PHCC and
KCuCl$_3$ and of the zig-zag chain, a model of theoretical interest in
its incommensurate regime. The extent of the incommensurate regime in
the zig-zag chain is satisfactorily obtained in an analytical
approximation close to the dimer point and this result provides simple
quantitative predictions for possible candidates for this interesting
phase. The magnon spectra of PHCC and KCuCl$_3$ as measured in
inelastic neutron scattering experiments can be used to determine the
exchange energies between the individual spins forming the basic
dimers by applying the dimer expansion to fourth, resp. sixth
order. KCuCl$_3$ turns out to be appropriately described as a system
of interacting chainlike structures with alternating couplings rather
than ladders. For PHCC a consistent set of exchange energies is
presented for interdimer interactions between dimers which are nearest
neighbours only. For the material TlCuCl$_3$, which is similar to, but
stronger interacting than KCuCl$_3$, the application of the dimer
expansion approach to the magnon spectra will be presented together
with new experimental results \cite{OKTKMM01}.
\begin{table}[ht]
\caption{Considered interaction constants of KCuCl$_3$ and PHCC. The energy scale is given by $J_{(000)}$.}
\label{tab:1}
\begin{tabular}{lllll}
\hline\noalign{\smallskip}
        &\multicolumn{2}{c}{KCuCl$_3$}&\multicolumn{2}{c}{PHCC} \\
\noalign{\smallskip}\hline\noalign{\smallskip}
$(nlm)$ & $(ij)$ & $J_{(nlm)}^{(ij)}$ & $(ij)$ & $J_{(nlm)}^{(ij)}$ \\
\noalign{\smallskip}\hline\noalign{\smallskip}
$(100)$                    & $(11),(22)$ & $0.000$   & $(11),(22)$ & 0.32 \\
                           & $(12)$      & 0.100     & $(12),(21)$ & 0.08 \\
$(201)$                    & $(21)$      & 0.188     & -           & -     \\
$(1\pm\frac 1 2 \frac 1 2)$& $(11),(22)$ & 0.200     & -           & -     \\
                           & $(12),(21)$ & 0.040     & -           & -     \\
$(001)$                    & -           & -         & $(11),(22)$ & 0.13 \\
                           & -           & -         & $(12)$      & 0.00 \\
$(101)$                    & -           & -         & $(12)$      & 0.09 \\
$(10-\!\!1)$               & -           & -         & $(21)$      & 0.09 \\
\noalign{\smallskip}\hline\noalign{\smallskip}
$(000)$                    &             & 4.250 meV &             & 2.05 meV \\
\noalign{\smallskip}\hline
\end{tabular}
%\vspace*{5cm}  % with the correct table height
\end{table}

\end{document}